\documentstyle[12pt,epsf]{article}

\def\hybrid{\topmargin -20pt    \oddsidemargin 0pt
        \headheight 0pt \headsep 0pt
        \textwidth 6.25in       
        \textheight 9.5in       
        \marginparwidth .875in
        \parskip 5pt plus 1pt   \jot = 1.5ex}

\hybrid

\def\baselinestretch{1.2}

\catcode`\@=11

\def\marginnote#1{}
\newcount\hour
\newcount\minute
\newtoks\amorpm
\hour=\time\divide\hour by60
\minute=\time{\multiply\hour by60 \global\advance\minute by-\hour}
\edef\standardtime{{\ifnum\hour<12 \global\amorpm={am}%
        \else\global\amorpm={pm}\advance\hour by-12 \fi
        \ifnum\hour=0 \hour=12 \fi
        \number\hour:\ifnum\minute<10 0\fi\number\minute\the\amorpm}}
\edef\militarytime{\number\hour:\ifnum\minute<10 0\fi\number\minute}

\def\draftlabel#1{{\@bsphack\if@filesw {\let\thepage\relax
   \xdef\@gtempa{\write\@auxout{\string
      \newlabel{#1}{{\@currentlabel}{\thepage}}}}}\@gtempa
   \if@nobreak \ifvmode\nobreak\fi\fi\fi\@esphack}
        \gdef\@eqnlabel{#1}}
\def\@eqnlabel{}
\def\@vacuum{}
\def\draftmarginnote#1{\marginpar{\raggedright\scriptsize\tt#1}}

\def\draft{\oddsidemargin -.5truein
        \def\@oddfoot{\sl preliminary draft \hfil
        \rm\thepage\hfil\sl\today\quad\militarytime}
        \let\@evenfoot\@oddfoot \overfullrule 3pt
        \let\label=\draftlabel
        \let\marginnote=\draftmarginnote
   \def\@eqnnum{(\theequation)\rlap{\kern\marginparsep\tt\@eqnlabel}%
\global\let\@eqnlabel\@vacuum}  }

\def\preprint{\twocolumn\sloppy\flushbottom\parindent 2em
        \leftmargini 2em\leftmarginv .5em\leftmarginvi .5em
        \oddsidemargin -.5in    \evensidemargin -.5in
        \columnsep .4in \footheight 0pt
        \textwidth 10.in        \topmargin  -.4in
        \headheight 12pt \topskip .4in
        \textheight 6.9in \footskip 0pt
        \def\@oddhead{\thepage\hfil\addtocounter{page}{1}\thepage}
        \let\@evenhead\@oddhead \def\@oddfoot{} \def\@evenfoot{} }

\def\numberbysection{\@addtoreset{equation}{section}
        \def\theequation{\thesection.\arabic{equation}}}

\def\underline#1{\relax\ifmmode\@@underline#1\else
        $\@@underline{\hbox{#1}}$\relax\fi}

\def\titlepage{\@restonecolfalse\if@twocolumn\@restonecoltrue\onecolumn
     \else \newpage \fi \thispagestyle{empty}\c@page\z@
        \def\thefootnote{\fnsymbol{footnote}} }

\def\endtitlepage{\if@restonecol\twocolumn \else \newpage \fi
        \def\thefootnote{\arabic{footnote}}
        \setcounter{footnote}{0}}  

\catcode`@=12
\relax

\def\figcap{\section*{Figure Captions\markboth
        {FIGURECAPTIONS}{FIGURECAPTIONS}}\list
        {Figure \arabic{enumi}:\hfill}{\settowidth\labelwidth{Figure
999:}
        \leftmargin\labelwidth
        \advance\leftmargin\labelsep\usecounter{enumi}}}
 \relax
\def\tablecap{\section*{Table Captions\markboth
        {TABLECAPTIONS}{TABLECAPTIONS}}\list
        {Table \arabic{enumi}:\hfill}{\settowidth\labelwidth{Table
999:}
        \leftmargin\labelwidth
        \advance\leftmargin\labelsep\usecounter{enumi}}}
 \relax
\def\reflist{\section*{References\markboth
        {REFLIST}{REFLIST}}\list
        {[\arabic{enumi}]\hfill}{\settowidth\labelwidth{[999]}
        \leftmargin\labelwidth
        \advance\leftmargin\labelsep\usecounter{enumi}}}
 \relax
\makeatletter
\newcounter{pubctr}
\def\publist{\@ifnextchar[{\@publist}{\@@publist}}
\def\@publist[#1]{\list
        {[\arabic{pubctr}]\hfill}{\settowidth\labelwidth{[999]}
        \leftmargin\labelwidth
        \advance\leftmargin\labelsep
        \@nmbrlisttrue\def\@listctr{pubctr}
        \setcounter{pubctr}{#1}\addtocounter{pubctr}{-1}}}
\def\@@publist{\list
        {[\arabic{pubctr}]\hfill}{\settowidth\labelwidth{[999]}
        \leftmargin\labelwidth
        \advance\leftmargin\labelsep
        \@nmbrlisttrue\def\@listctr{pubctr}}}
 \relax
\makeatother
\newskip\humongous \humongous=0pt plus 1000pt minus 1000pt

\newif\ifdtup

\relax

\def\be{\begin{eqnarray}}
\def\ee{\end{eqnarray}}
\def\ba{\begin{eqnarray}}
\def\ea{\end{eqnarray}}

\def\D{\Delta}

\def\A{\Alpha}

\def\d{\partial}
\def\D{\Delta}

\def\p{\pi}

\def\m{\mu}
\def\n{\nu}

\def\L{\Lambda}

\def\no{\noindent}

\def\IR{\relax{\rm I\kern-.18em R}}

\def \A { {\bar A} }

\def\diag{{\rm diag}}

\def\IR{\relax{\rm I\kern-.18em R}}
\def\inv{^{\raise.15ex\hbox{${\scriptscriptstyle -}$}\kern-.05em 1}}

\def\tL{{\tilde L}}
\def\L{\Lambda}

\begin{document}

\renewcommand{\theequation}{\arabic{equation}}

\newcommand{\beq}{\begin{equation}}
\newcommand{\eeq}[1]{\label{#1}\end{equation}}
\newcommand{\ber}{\begin{eqnarray}}
\newcommand{\eer}[1]{\label{#1}\end{eqnarray}}
\newcommand{\eqn}[1]{(\ref{#1})}
\begin{titlepage}
\begin{center}

\hfill NTUA-75/99\\ \hfill hep--th/9911134\\

\vskip 1.2in

{\large \bf On Non-Compact Compactifications with Brane Worlds}
\vskip 0.6in

{\bf A. Kehagias}
\vskip 0.1in
{\em Physics Dept. National Technical University, \\
 157 73 Zografou, Athens, Greece    \\
{\tt kehagias@mail.cern.ch}}\\
\vskip .2in

\end{center}

\vskip .6in

\centerline{\bf Abstract }

\no The possibility of neutral, brane-like solutions in a higher
dimensional setting is discussed. In particular, we describe  a
supersymmetric solution in six dimensions which can be interpreted
as a ``three-brane"  with a non-compact transverse space of finite
volume. The construction can be generalized to
$n\!+\!4$-dimensions and the result is a $n\!+\!3$-brane
compactified on a $n-1$-dimensional Einstein manifold with a
non-compact extra dimension. We find that there always exists  a
massless graviton trapped in four-dimensions while a bulk abelian
gauge field gives rise to a unique four-dimensional massless
photon.
Moreover,  all massless modes are accompanied   by  massive KK
states and we show that it is possible  in such a scenario the
masses of the KK states to be at the TeV scale without
hierarchically large extra dimensions.

\vskip .1cm
\no

\vskip 2cm \noindent NTUA-75/99\\ November 1999\\
\end{titlepage}
\vfill
\eject

\def\baselinestretch{1.2}
\baselineskip 16 pt
\noindent

\def\tT{{\tilde T}}
\def\tg{{\tilde g}}
\def\tL{{\tilde L}}


\section{Introduction}

There is a renewal interest in higher-dimensional theories after
the realization that string scale can be lowered even  to few {\rm
TeV}s \cite{LB}. In this case, extra dimensions are expected to
open up at this scale  as was originally proposed in \cite{IA}
motivated by the scale of supersymmetry breaking. It has also been
realized \cite{ADD}, that the size of  some dimensions can even
macroscopic as long as the Standard Model (SM) sector lives solely
in four-dimensions, in a three-brane for example. This proposal
offers also a natural explanation  to  a long-standing problem in
all efforts to extend the SM, namely, the hierarchy problem. The
latter translates into the fact that the ratio $m_{EW}/M_{Pl}$ of
the electroweak scale $m_{EW}\sim 10^3 \ {\rm GeV}$ to the Planck
scale $M_{Pl}= G_N^{1/2}\sim 10^{18}\ {\rm GeV}$ is unnatural very
small.
 According to this proposal, the hierarchy is due to the geometry
 of the higher-dimensional space-time. In particular,
  the higher $4+n$ dimensional theory with $n\geq 2$ has a $4+n$
dimensional  {\rm TeV} scale Planck mass $M_{Pl(4+n)}$  while the
scale $R_c$ of the extra $n$ dimensions is less than a millimeter.
The proposal has been designed in such a way as to generate the
hierarchy $m_{EW}/M_{Pl}$.

These ideas have been pushed further by modeling four-dimensional
space-time as a brane embedded  in higher dimensions which  is
reminiscent of some old proposals that tried to model our world as
a ``domain wall''\cite{RS}. This is realized in modern terms by
the brane world where our universe is viewed  as a three-brane (or
p-brane with appropriate compactified directions, or intersecting
branes, etc.). An early example  was the Horava-Witten picture for
the non-perturbative heterotic $E_8\times E_8$ string \cite{hw}
and its relevance for the construction of realistic
phenomenological models \cite{wit}. A general feature of all these
models is that  the gauge sector lives on the brane whereas
gravity propagates in the bulk. Although a bulk SM (or partial
bulk) has some nice properties like power-law running couplings
\cite{TV},\cite{DDG} and consequently a lower  unification scale
\cite{DDG}, by trying to put the gauge sector in the bulk in
models with large extra dimensions, for example, one faces the
problem that Kaluza-Klein (KK) states of the photons should have
already been observed. Thus, a bulk SM constraints  the size of
the extra dimensions to be more than around 1{\rm TeV} depending
on their number \cite{NY}. Thus, it seems that a bulk gauge sector
is in conflict with large extra-dimensions.

Here, we will consider classical supergravity solutions which can
be interpreted as ``domain walls'' in higher dimensions. The
domain walls we will construct  are formed in vacuum, contrary to
the usual cases where other fields are present, usually scalars.
So their formation is due to gravity.  In particular, out of the
domain wall, space-time is Ricci-flat  and all the energy is
concentrated at the position of the wall. Thus, our model differs
essentially from other similar proposals. In the model of Randall
and Sundrum (RS) for example \cite{RS1}, there is a bulk
cosmological constant while in that of Cohen and Kaplan scalar
fields \cite{CK}. In these models a massless four-dimensional
graviton appears on the domain wall due to the finiteness of the
transverse space as in the old KK literature \cite{GZ}. In our
case, there is no cosmological constant or matter fields  and
space-time is Ricci flat (or even completely flat) everywhere
except at the position of the wall.  Thus, it can also be
considered as a vacuum in heterotic theories. In the latter case
there exist bulk gauge bosons and one may also ask if there exist
massless gauge bosons trapped, like the graviton, in the wall. We
find that indeed both bulk graviton and photons give rise to a
four-dimensional massless graviton and photons with a discrete
spectrum of KK modes. Gauge bosons in the RS case have been
discussed in \cite{PR}.  We also show that in this scenario it is
possible for a {\rm TeV} scale internal space.

\section{Novel vacua with non-compact dimensions}

One way of constructing string theory vacua is to look for
classical supergravity solutions. As supergravity is the
low-energy limit of string theory, supergravity solutions describe
accordingly low-energy string vacua. There will be
$\alpha'$-corrections as well as string-loop corrections to these
solutions but, nevertheless, these solutions will still be valid
in some appropriate limits. These vacua are constructed by solving
the classical field equations with appropriate fields turned on.
Usually such fields are antisymmetric p-forms as well as scalars
like the dilaton, axion etc.,   which appear in almost all
supergravity theories. Some of these classical solutions are
interpreted  as the D-branes of string theory. There are also
string vacua where the are no other fields, except the graviton,
turned on.  Such vacua are necessarily Ricci-flat
\be
R_{MN}=0\, , \label{Einstein} \ee i.e., they satisfy  vacuum
Einstein equations. Solutions to the above equations with
four-dimensional Poincar\'e invariance are provided by
$M^4\!\times\! X$ where $M^4$ is ordinary Minkowski space-time and
$X$ is a Ricci-flat manifold.  Supersymmetry demands that $X$
should be a manifold  of $U(1)^6, ~SU(2)\times U(1)^2$ or $SU(3)$
holonomy. Manifolds with such holonomies can be either compact or
non-compact. We recall for example the case of the compact $K3$
surface and the non-compact
 Eguchi-Hanson gravitational instanton,   both of $SU(2)$ holonomy.
  The four-dimensional Plank mass $M_{P}$ is proportional to the
volume $V(X)$ of $X$
\be
M_{P}^2= M_s^8\,  V(X)\, , \label{Mp} \ee where $M_s^2\sim
1/\alpha'$ is the string-mass scale. Propagating gravity therefore
exists in four dimensions  if the volume of $X$ is finite. In this
case, $X$ must be compact so that $X$ is either $T^6,
~K3\!\times\!T^2$ or $CY_3$ depending on the number of surviving
supersymmetries. It should be stressed, however, that there are
also non-compact spaces of finite volume which, according to
eq.(\ref{Mp}) will lead to a four-dimensional dynamical gravity.
Such spaces have been considered in the Kaluza-Klein programme
\cite{GZ} and discussed \cite{wet} in connection to  the
chiral-fermion problem \cite{WWW}. The drawback of non-compact
spaces of finite volume is that they suffer from singularities.
However, although singularities are considered in general to be
disastrous, there exist singularities which are quite mild in the
sense that they can be attributed to some  form of matter. These
are delta-function singularities which may be interpreted as
strings, domain walls or fundamental branes in general.
Supergravity solutions with a bulk cosmological constant and such
singularities have been constructed in \cite{KK} Our aim here is
to solve eq.(\ref{Einstein}) with a non-compact internal space $X$
of finite volume and delta-functions singularities.

In five dimensions the only solution to eq.(\ref{Einstein}) with
four-dimensional Poincar\'e symmetry $ISO(1,3)$ is flat space i.e.,
$ISO(1,3)$-invariance and Ricci-flatness
implies flat space in five dimensions.
However, this is not true in higher dimensions. For example let us try
to solve eq.(\ref{Einstein}) in six-dimensions
with the $ISO(1,3)\!\times\!U(1)$-invariant metric
\be
ds^2=-dt^2+dx_1^2+dx_2^2+dx^2_3+e^{2\L(z)}\left(dz^2+d\phi^2\right)\,
, \label{met1} \ee where $0<\phi\leq2\p a$ parametrize an $S^1$
with radius $a$ and $-\infty<z<\infty$. Eq.(\ref{Einstein}) is
then written as
\be
R_{zz}=R_{\phi\phi}=-\L''=0 \, . \label{e}
\ee
The solution of eq.(\ref{e}) for $\L$ is, up to an irrelevant constant,
\be
\L=\L(z)=\epsilon\, \mu\, z\, ,  ~~\epsilon =\pm 1\, , \ee where
$\mu>0$ is a dimensionfull constant. The metric (\ref{met1}) turns
out then to be
\be
ds^2=\eta_{\m\n}dx^\m dx^\n+e^{2\epsilon \mu
z}\left(dz^2+d\phi^2\right)\, , ~~~~\m,\n=0,1,2,3 \ee with
$\eta_{\mu\nu}=\diag(-1,1,1,1)$  the Minkowski metric. One
observes that the metric above is locally flat since it is an
alternative  way of writing the metric of a flat six-dimensional
Minkowski space. Indeed, the transformation $\mu z=
\epsilon\,\ln(\mu r)$ gives  a manifestly flat-form of the metric
(for $\mu=1/a$). For $\mu\neq 1/a$, we get as solution the product
of
 four-dimensional Minkowski space
with a  two-dimensional flat  cone since in this case
there exist a deficit angle
$2\pi(1-a\mu)$ for $\phi$.

In solving eq.(\ref{e}), we have made  the assumption that both
$\L$ and its first derivative are everywhere continuous. However,
by relaxing the continuity conditions we may get other solutions
as well. For example, there exist    piecewise flat solutions in
which $\L$ is   continuous but with discontinuous first
derivatives. Such solutions are provided  by the choice
\be
\L(z)=\epsilon\,\mu \, |z|\, ,  ~~\epsilon =\pm 1\, . \label{L}
\ee
As a result, we get
\be
\L''=2\,\epsilon\,\mu\, \delta(z)\, ,
\ee
and the metric  turns out to be
\be
ds^2=\eta_{\m\n}dx^\m dx^\n+
e^{2 \epsilon \mu |z|}\left(dz^2+d\phi^2\right)\, . \label{met0}
\ee
The sing $\epsilon$ will be determined in a moment after we calculate the
energy-momentum tensor.
The Ricci tensor develops delta-function singularities. Indeed,
from eq.(\ref{e}) we find that
\be
R_{zz}=R_{\phi\phi}=-2\,\epsilon\,\mu\, \delta(z) \, , ~~~
R=-4\epsilon\,\mu \delta(z)e^{-2\epsilon\mu  |z|}\, , \label{ee}
\ee where $R$ is the scalar curvature. Thus, the metric
(\ref{met0}) is again everywhere  flat for $\mu=1/a$ (or locally
flat for   $\mu\neq 1/a$ but now it develops a delta-function
singularity at the point $z=0$. To see if this singularity can be
attributed to some form of matter, we have to calculate the
energy-momentum tensor. The latter may be read off from
\be
T_{MN}=
\frac{1}{8\pi G_6}\left(R_{MN}-\frac{1}{2}G_{MN}R\right)\, , \label{EM}
\ee
where $G_n$ is, in general, the $n$-dimensional Newton constant.
By using eq.(\ref{ee})  we get that
\begin{eqnarray}
T_{\mu\nu}&=&
{1\over 4\pi G_6}\epsilon\mu e^{-2\epsilon \mu |z|}\delta(z)\eta_{\mu\nu}
\nonumber \\
T_{zz}&=&T_{\phi\phi}=0\, .
\end{eqnarray}
Positivity now of the energy-density,
$$T_{00}=\rho=-\epsilon{\mu\over 4\pi G_6} e^{-2\epsilon \mu
|z|}\delta(z)$$ demands  that $\epsilon=-1$. As a result, the
metric turns out to be
\be
 ds^2=\eta_{\m\n}dx^\m dx^\n+
e^{-2 \mu |z|}\left(dz^2+d\phi^2\right)\, . \label{met}
\ee
This metric describes a ``string''
(string if we count the codimension of the object, or a three-brane if we
count its actual dimensions) with a four-dimensional
world-volume  in
six dimensions as can be seen from the  form of energy-momentum tensor
\be
T_{MN}=\rho~ \diag(1,-1,-1,-1,0,0)\, . \ee It should be stressed
that this three-brane is not the one of type IIB theory since it
is neutral and, in particular, it does not carry any RR charge to
justify its name. However, it has a four-dimensional world-volume
and for this reason we will call it three-brane. The transverse
space of this three-brane  is a non-compact surface $\Sigma_2$
with metric
\be
ds_{\bot}= e^{-2 \mu |z|}\left(dz^2+d\phi^2\right)\, . \label{mm}
\ee Remarkably, the area of the surface $\Sigma_2$  is
\be
V_{\bot}=2\pi a\int_{-\infty}^{\infty}e^{-2\mu  |z|}\,
dz=2\pi\mu^{-1}a  <\infty\, , \label{mmm}\ee and thus, a massless
graviton is expected in four-dimensions as will see later.

\subsection{Supersymmetry}

 We will examine now if the background of eq.(\ref{met}) is
supersymmetric. In this case, the gravitino shifts
\be
\delta\psi_M=D_M\epsilon\, , \label{grav} \ee where
$D_M=\d_M+\omega_{MAB}\Gamma^{AB}/4$ is the spin connection,
 vanish for appropriate spinors $\epsilon$. We may split $\epsilon$  as
\be
\epsilon=\theta\otimes \eta\, ,
\ee
where $\theta,\, \eta$ are  four-dimensional and two-dimensional spinors,
respectively. We choose for the gamma matrices the representation
\be
\Gamma^\alpha&=&\gamma^\alpha\otimes 1\, , ~~ \alpha=0,1,2,3,
\nonumber \\ \Gamma^z&=& \gamma^5\otimes \sigma^1\, , ~~~
\Gamma^\phi= \gamma^5\otimes \sigma^2\, , \label{gamma} \ee where
$\gamma^\alpha$ are four-dimensional gamma matrices and
 $\sigma^{1,2}$ are
Pauli matrices.
The  vanishing of the gravitino shifts is then  equivalent to the
existence of covariantly constant spinors in the transverse
space  $\Sigma_2$ with metric (\ref{mm}). In particular, the number of
supersymmetries in four-dimensions is the number of independent
Killing spinors in the transverse space $\Sigma_2$.
The Killing spinor equation
is
\be
D_i\eta=0\, , \ee where $D_i=\d_i+\omega_{iab}\sigma^{ab}/4$ is
the spin connection in the two-dimensional space $\Sigma_2$. For
the metric (\ref{mm}) the Killing spinor equation splits as
follows
\be
&&\d_z\eta=0\, , \label{k1}\\ &&\left(\d_\phi+{1\over
2}\L'\sigma^2\sigma^1\right)\eta=0\, . \label{k2} \ee The solution
to  eq.(\ref{k2}) for  $\eta$ is given by
\be
\eta=\left(\Theta(-z)\Big{(}\sigma^2\cos{\mu\over 2}\phi+
\sigma^1\sin{\mu\over 2}\phi\Big{)}+\Theta(z)\Big{(} \cos{\mu\over
2}\phi-\sigma^1\sigma^2\sin{\mu\over 2}\phi\Big{)}\right) \eta_0\,
, \label{k3} \ee where $\Theta(z)=0,1$ for $z<0,\,z>0$,
respectively is the step function and $\eta_0$ is a two-component
constant spinor. Substituting eq.(\ref{k3}) back in eq.(\ref{k1}),
we get
\be
\d_z\eta=\delta(z)\left(\phantom{{\mu\over
2}}\!\!\!\!\!1-\sigma^2\right)\left(\cos{\mu\over 2}\phi
-\sigma^1\sin{\mu\over 2}\phi\right) \eta_0\, . \ee The matrix
$(1-\sigma^2)$ has one zero eigenvalue and thus, there exist only
one covariant constant  spinor
 localized at $z=0$.
  Therefore, the solution we found breaks half of the
supersymmetries. The non-zero eigenvalue of $\d_z\eta$ is the
Golstone fermion and
 lives only inside the wall due to the delta function.

\section{Domain Walls in higer dimensions}

We have seen  that the conditions of four-dimensional Poincar\'e
invariance and Ricci-flatness lead, in  six dimensions, to
three-brane like solutions. This construction can be generalized
to higher dimensions. Here we will consider an n+4-dimensional
space-time of the form $M^{1,3}\times X^n$ with metric
\be
ds^2=-dt^2+dx_1^2+dx_2^2+dx_3^2+e^{2\L(z)}\left(dz^2+
d\sigma^2\right)
\, ,
\label{met3}
\ee
where
\be
d\sigma=k_{ij}(y)dy^idy^j\, ,
\ee
is the metric of an $n\!-\!1$-dimensional space $\Sigma$.
The Ricci tensor for the
metric (\ref{met3}) is
\begin{eqnarray}
R_{\mu\nu}&=&0\, , \nonumber \\
R_{zz}&=&-(n-1)\L''\, , \nonumber \\
R_{ij}&=& R(k)_{ij}-\L''k_{ij}-(n-2)\L'^2k_{ij}\, , \label{r}
\end{eqnarray}
where $R(k)_{ij}$ is the Ricci tensor of the space $\Sigma$.
Eq.(\ref{Einstein}) is then satisfied by
\begin{eqnarray}
\L&=&\epsilon\, \mu\,  z\, , ~~\epsilon =\pm 1\, , \nonumber \\
R(k)_{ij}&=&(n-2)\mu^2 k_{ij}\, ,
\end{eqnarray}
so that the space $\Sigma$ is a Einstein space of positive
constant scalar curvature $(n\!-\!1)(n\!-\!2)\mu^2$. $\Sigma$ can
be any compact Einstein space and, in particular, the solution is
flat n+4-dimensional space-time if $\Sigma$ is the round sphere
$S^{n-1}$ with radius $1/\mu$ .

As in the six-dimensional case we discussed before, we may take
$\L$ to be continuous but with discontinuous first derivatives. In
this case  $\L$ is given by eq.(\ref{L}) and the metric turns out
to be
\be
ds^2=\eta_{\m\n}dx^\m dx^\n+e^{2\epsilon\mu |z|}
\left(dz^2+k_{ij}dy^idy^j\right)\, .
\label{met2}
\ee
Here again, the space is everywhere Ricci-flat
except at the point $z=0$ where
it develops a delta-function singularity such that
\be
R_{\mu\nu}=0\, , ~~~ R_{ij}=-2\,\epsilon\,\mu\,\delta(z)\, k_{ij}\, ,
~~~ R_{zz}=-2(n-1)\,\epsilon\,\mu\,\delta(z)\, .
\ee
The energy-momentum tensor can be calculated  from eq.(\ref{EM}) and
we find
\begin{eqnarray}
T_{\mu\nu}&=&\epsilon(n-1){\mu\over 4\pi G_{(n+4)}}
e^{-2\epsilon \mu |z|}\delta(z)\eta_{\mu\nu}
\nonumber \\
T_{ij}&=& \epsilon(n-2){\mu\over 4\pi G_{(n+4)}}
e^{-2\epsilon \mu |z|}\delta(z)k_{ij}\, , \nonumber \\
T_{zz}&=&0\, .
\end{eqnarray}
Positivity of the energy-density requires again $\epsilon=-1$ and
from the form of the energy-momentum tensor we see that the
solution
\be
ds^2=\eta_{\m\n}dx^\m dx^\n+e^{-2\mu |z|}
\left(dz^2+k_{ij}dy^idy^j\right)\, , \label{met2} \ee represents a
domain wall at $z=0$. This domain wall however,  is not a flat
$n+3$-dimensional Minkowski space-time $M^{1,n+2}$ as in the usual
case but rather  is of $M^{1,3}\!\times\!\Sigma$ topology. In a
sense it may be viewed as a compactified flat Minkowski space-time
on a compact Einstein space $\Sigma$ as in the old Kaluza-Klein
programme. The transverse space $X^n$ to the four-dimensional
Minkowski space-time has metric
\be
ds^2=e^{-2\mu |z|}
\left(dz^2+k_{ij}dy^idy^j\right)\, ,
\label{met4}
\ee
and its volume is
\be
V_{\bot}=V(\Sigma)\int_{-\infty}^\infty e^{-n\mu|z|}dz={2\over n}
V(\Sigma)\mu^{-1}<\infty\, ,
\ee
which is again finite.


\section{The bosonic spectrum}

We will examine now the spectrum of small fluctuations of the bulk
fields. As usual, the bulk fields are the graviton, scalars (like
dilaton or axions), gauge fields, antisymmetric tensor fields,
fermions and gravitinos.  We will discuss here  the case of bulk
graviton and gauge fields.

To study the spectrum, we need certain Hodge-de Rham operators
$\D_p$ in the $n$-dimensional space $X^n$. The action of the
latter on scalars $Y$ and one-forms $Y_m$ is
\be
\D_0Y&=& -\nabla_m\nabla^mY\, , \nonumber \\ \D_1Y_m&=&
-\nabla_p\nabla^pY_m+{R_m}^pY_p\, , ~~~~~m,k,p,q=1,...,n
\, . \ee
 Another operator which is involved in the discussion is the
 Lichnerowitz operator $\D_L$ which acts
on traceless transverse symmetric two-tensors as
\be
\D_Lh_{mk} = -\nabla_p\nabla^p h_{mk}+{R_m}^ph_{kp}+{R_k}^ph_{mp}
-2R_{mpkq}h^{pq}\, . \label{Li} \ee Next we need the eigenvalues
of the Hodge-de Rham and Lichnerowitz
 operators $\D_p$ and $\D_L$, respectively on the space $X$. We will work out explicitly
 the eigenvalue problem for the Laplace operator $\D_0$ which is directly involved in the
 discussion for massless four-dimensional gravitons, whereas for the rest, we will
 find bounds on their lowest eigenvalue.
 For simplicity, we will assume that the metric of $X$ is
\be
ds^2_\bot= e^{-2\mu|z|}\left(dz^2+{1\over
\mu^2}d\Omega_{n-1}^2\right)\, , \ee where $d\Omega_{n-1}^2$ is
the metric on the unit $n\!-\!1$-sphere $S^{n-1}$. Thus, away from
the $z=0$ point, the transverse space is  flat $n$-dimensional
Euclidean space and all the curvature is at the $z=0$ point.

We will first consider the scalar Laplacian $\D_0$ and its eingenvalue problem
\be
\D_0Y=M^2Y\, . \label{ein0}
\ee
By writing
\be
Y_\ell(y^i,z)=e^{(n-2)\mu|z|/2}Z(z)\Phi_\ell(y^i)\, , \ee where
$\Phi_\ell(y^i)$ are the eigenfunctions of the scalar Laplacian
$\D_0(S^{n-1})$ on the unit  $S^{n-1}$ with eigenvalues
$\ell(\ell+n-1)$
\be
\D_0(S^{n-1})= \ell(\ell\!+\!n\!-\!2) \Phi_\ell\, ,
~~~\ell=0,1,... \ee we get from eq.(\ref{ein0}) that  $Z(z)$
satisfies
\be
-{d^2\over dz^2}Z+\left({(n-2)^2\over
4}\mu^2+\ell(\ell\!+\!n\!-\!2)\mu^2-(n-2) \mu\delta(z)
\right)Z=M^2e^{-2\mu |z|}Z      \, . \label{sl} \ee The problem
has been reduced to an one-dimensional Schr\"odinger equation with
potential
\be
V(z)=\left({(n\!-\!2)^2\over 4}+\ell(\ell\!+\!n\!-\!2)\right)\mu^2
-\!(n\!-\!2)\, \mu\, \delta(z)\, . \label{pot} \ee In general, an
attractive potential $V(z)=-g^2\delta(z)$ supports a single bound
state of energy $-g^2/4$. We see that this bound state satisfies
eq.(\ref{sl}) for $\ell=0$ and $M=0$. As a result, the
four-dimensional massless graviton is just the unique bound state
of the potential (\ref{pot}). To find the rest of the spectrum, we
have to solve eq.(\ref{sl}) with appropriate boundary conditions.
If we denote by $Z_+(z),Z_-(z)$ the solution for $z>0,z<0$,
respectively, the boundary conditions are
\be
Z_+(0)=Z_-(0)\!&=&\!Z(0)\, , \\ \label{conti}
Z'_+(0)-Z'_-(0)\!&=&-\mu (n-2)Z(0)\, , \label{d} \ee where prime
denotes differentiation with respect to $z$('=$d/dz)$. In
addition, there exist one more relation $Z$ has to satisfy,
namely,
\be
e^{-(n-2)\mu z}Z_+(z)Z'_+(z)|_{z=\infty}-
e^{(n-2)\mu z}Z_-(z)Z'_-(z)|_{z=-\infty}=0\, , \label{inf}
\ee
which is just
the condition of conservation of the current $J^p=Z\d^p Z$
on the transverse space, i.e.,
\be
\int dzd^{n\!-\!1}y\,
\d_p\left(e^{-(n-2)\mu |z|}\sqrt{h}Z\d^pZ \right)\, .
\ee
Continuity of $Z$ eq.(\ref{conti}) and the condition eq.(\ref{inf}),
specify the solutions to be, up to a multiplicative constants,
\be
Z_+(z)&=&M^{(2-n)/ 2} J_\nu\Big{(}{M\over \mu}e^{-\mu z}\Big{)}\,
, \nonumber \\ Z_-(z)&=&M^{(2-n)/ 2}J_\nu\Big{(}{M\over \mu}e^{\mu
z}\Big{)}\, , \label{Z}
\ee where $J_\nu$ are the Bessel functions
with
\be
\nu={2\ell+n-2\over 2}\, . \label{n} \ee The factor $M^{(2-n)/ 2}$
in eq.(\ref{Z}) has been inserted in order  the limit $M\to 0$ to
give the zero-mode eigenfunction
\be
Z_{0}(z)\sim exp^{-(n-2)\mu|z|}\, . \label{Z0}\ee
 The other solution to
eq.(\ref{sl}) the second Bessel function
 $Y_\nu$ fails
to satisfy eq.(\ref{inf}). Finally, from the last condition eq.(\ref{d}),
we get
\be
{M\over \mu} J_\nu'\Big{(}{M\over \mu}\Big{)}=
{n-2\over 2} J_\nu\Big{(}{M\over \mu}\Big{)}\, ,
\ee
which can be written, after using Bessel-function identities, as
\be
{M\over \mu} J_{\ell+n/2}\Big{(}{M\over \mu}\Big{)}= \ell
J_{\ell-1+n/2}\Big{(}{M\over \mu}\Big{)}\, . \label{jj} \ee Thus,
the spectrum is $M_{k,\ell}$ where $M_{k,\ell}$ satisfies
eq.(\ref{jj}). In particular, for $\ell=0$, we find that $M_{k,0}$
is
\be
M_{k,0}=\mu j_{n/2}^{(k)}\, , ~~~ M_{0,0}=0\, , ~~~~~~k=1,2...\, ,
\label{sp} \ee where $j_{n/2}^{k}$ are the zeros of $J_{n/2}$.
Note that $M_{0,0}=0$ corresponds to the bound state we found
before. It should be noted that for $\ell\neq 0$, the value $M=0$
which solves eq.(\ref{jj}), gives $Z(z)=0$ and thus, there exist
only zero eigenvalue $M_{0,0}$ with corresponding eigenfunction
(\ref{Z0}). The spectrum of $\D_0$ is given in table (\ref{t1})
\be
\begin{tabular}{|c|c|}\hline
eigenvalues of $\D_0$ & dim of $SO(n-1)$\\ \hline  \hline
$j_{n/2}^{(k)}$& 1\\ \hline $\phantom{{X{X\over X}\over
X}}M_{k,\ell}\,, \ell\neq 0$&${(2\ell+n-2)(\ell+n-3)!\over
(n-2)!\ell!}$\\ \hline
\end{tabular}\,  \label{t1}
\ee

In particular, for the $n=2$ case, where the transverse space is
$\Sigma_2$ with metric (\ref{mm}), eq.(\ref{sl}) is written as
\be
-{d^2\over dz^2}Z+\ell^2\mu^2Z=M^2e^{-2\mu |z|}Z \, . \label{sll}
\ee The solution is then
\be
Z_+(z)= J_\ell\Big{(}{M\over \mu}e^{-\mu z}\Big{)}\,, ~~~~
Z_-(z)=J_\ell\Big{(}{M\over \mu}e^{\mu z}\Big{)}\, , \ee while the
eigenvalues (\ref{ein0}) are specified by
\be
{M\over \mu} J_{\ell+1}\Big{(}{M\over \mu}\Big{)}= \ell
J_{\ell}\Big{(}{M\over \mu}\Big{)}\, . \label{jjj} \ee The
$\ell=0$ tower consists of the zeroes of $J_1$ and the massless
mode is the $x=0$ of the equation $J_1(x)=0$. The rest of the
spectrum is obtained by solving eq.(\ref{jjj}).

 Concerning the operator $\D_1$, it is not
difficult to verify that its eigenvalues $M_1^2$ are strictly
positive, i.e., $M^2_1>0$ for $n\geq 2$. Indeed, from the
eigenvalue problem
\be
-\nabla_p\nabla^p Y_m+R_{mp}Y^p=M_1^2Y_m\, , \label{m1} \ee we
see, by multiplying both sides with $Y_m^*$ and integrating over
$X$ that
\be
M_1^2\geq {1\over |Y_m^*Y^m|^2}\int R_{mp}Y^{*p}Y^m\, ,
~~~~|Y_m^*Y^m|^2=\int Y^{*}_mY^m \, . \ee Since now $R_{mp}$ is a
strictly positive matrix, we see that there is no zero eigenvalue.

\subsection{Graviton}

In order to find the  spectrum of small fluctuations  around the
background metric $g_{MN}$ of (\ref{met2}) we write
$\hat{g}_{MN}=g_{MN}+\delta g_{MN}$ and we keep only linear terms
in $ \delta g_{MN}=h_{MN}$ in the equation
\be
\hat{R}_{MN}(g+h)=0\, .\ee Then, we get that $h_{MN}(x^\mu,y^i,z)$
satisfies the equation
\be
\delta R_{MN}&=&-\nabla_K\nabla^K h_{MN}+{1\over 2}R_{MA}{h^A}_N+
{1\over 2}R_{NA}{h^A}_M-{R^A}_{MKN}{h^K}_A\nonumber \\
&&
+{1\over 2}\nabla_M\nabla^A h_{NA}
+{1\over 2}\nabla_N\nabla^A h_{MA}-{1\over 2}\nabla_M\nabla^M {h^A}_A\, ,
\label{h}
\ee
where $\nabla_M$
is the covariant derivative  with respect to the
background metric (\ref{met2}).  We may express the components of $h_{MN}$
as
\be
h_{\m\n}(x,y)&=&h_{\m\n}(x)Y(y)\, , \nonumber \\ h_{\m
n}(x,y)&=&B_\mu(x)Y_n(y)\, , \nonumber  \\ h_{mn}(x,y)&=&
V(x)Y_{mn}(y)+{1\over n}g_{mn}U(x)Y(y)\, , \label{exp} \ee where
$Y(y),Y_m(y)$ have been defined in eqs(\ref{ein0},\ref{m1}), and
$Y_{mn}(y)$ is transverse traceless.  We see from the expansion
(\ref{exp}) that we get in four dimensions a symmetric tensor
field $h_{\m\n}$ which contains the graviton, a vector $B_\mu$ and
the scalars $U,V$. We are particular interested for massless four
dimensional fields and we will examine if there are such massless
modes.
 Due to the invariance
\be
\delta g_{MN}=\nabla_M \xi_N+\nabla_N\xi_M\,
\ee
we may impose $n+4$ conditions on $h_{MN}$ which we choose to be
\be
\nabla^mh_{m\nu}=0\, , ~~~~\nabla^m h_{mn}={1 \over n}
g_{mn}\nabla^m {h^k}_k\, . \ee
 By using the expansion (\ref{exp}), we get
from the $(\mu\nu)$ component of eq.(\ref{h})
\be
&&\nabla_\rho\nabla^\rho h_{\mu\nu} -\nabla_\nu\nabla^\rho
{h^\rho}_\mu -\nabla_\nu\nabla^\rho {h^\rho}_\nu+
\nabla_\mu\nabla^\nu {h^\rho}_\rho=M^2h_{\mu\nu}\, , \label{hih}
\ee while from the $(\mu,n)$ components we get, among others,
\be
&&\nabla_\mu\nabla^\mu B_\nu=M_1^2 B_\nu\, ,  \label{bb}
\ee Eq.(\ref{hih}) is the equation for the four-dimensional
graviton while eq.(\ref{bb}) is the equation for the KK vector
$B_\mu$. Since $M^2$ is the eigenvalues of the scalar Laplacian in
$X^n$ which, as we have seen  has a zero mode,  a massless
graviton always exists and it is the unique bound state in the
attractive delta-function potential of eq.(\ref{pot}). The KK
modes of the four-dimensional graviton have masses given in table
(\ref{t1}). On the other hand, the KK vector $B_\mu$ is massive
since the operator $\D_1$ does not have a zero eigenvalue.
Proceeding as above for the scalars, we find that they are
massive. Thus, the only massless mode in four-dimensions of the
higher-dimensional bulk graviton  is the four-dimensional
graviton.


The massive KK modes of the graviton will affect the  Newton law
as usual generating Yukawa-type corrections \cite{ADD,Nwt,KS}. The
gravitational potential where also the massive KK modes of the
graviton are taken into account is given by \cite{KS}
\be
V(r)=-{1\over r}\sum_k d_k e^{-M_k r} \, , \ee where $d_k$ is the
degeneracy of the $k$-th massive KK state. Note that the
degeneracy is due to the $SO(n-1)$ invariance and the range is set
by the mass of the first KK state which is proportional to $\mu$.

\subsection{Gauge fields}

Let us now consider a $U(1)$ gauge field
$A_M(x,y^i,z)=(A_\mu,A_i,A_z)$ in the bulk geometry. We will
examine the spectrum which appears on the brane at $z=0$ due to
the bulk gauge field. The field equations for the gauge field is
just the Maxwell equations
\be
\nabla^MF_{MN}=0\, , \label{M} \ee where $\nabla_M$ is the
covariant derivative in the $n+4$-dimensional space-time and
$F_{MN}=\nabla_M A_N-\nabla_N A_M$ is the field strength.
Eq.(\ref{M}) follows from the action \be S_{n\!+\!1}=-{1\over
4g^2}\int d^{4\!+\!n}x F_{MN}F^{MN}\, . \label{action} \ee In
terms of the gauge field $A_M$ and in the covariant gauge,
eq.(\ref{M}) is written as
\be
-\nabla_M\nabla^MA_N +{R^M}_NA_M=0\, ,
~~~~\nabla^MA_M=0.\label{Ma} \ee By  writing the components of the
gauge field as
\be
A_\mu(x,y,z)=a_\mu(x)Y(y,z)\, , ~~~A_m(x,y,z)=a(x)Y_m(y,z)\, ,~~~
\ee and recalling eq.(\ref{r}) for the Ricci tensor in the
background (\ref{met2}), we get that $a_\mu,a$   satisfy
\be
\nabla^2_{(4)} a_\mu= M_0^2a_\mu\, , ~~~ \nabla^2_{(4)} a=
M_1^2a\, , \ee where $M_0^2,M_1^2$ are the eigenvalues of the
$\D_0,\D_1$ operators, respectively, i.e.,
\be
&&-\nabla^2_{(n)}Y=M_0^2Y\, , \label{o1}\\
&&-\nabla^2_{(n)}Y_i+{R_i}^jY_j=M_1^2Y_i\, ,\label{o2}
\ee Thus a massless four-dimensional photon exists if there exist
a zero mode of the Laplace operator in eq.(\ref{o1})  while a
massless four-dimensional scalar appears if the operator in
eq.(\ref{o2}) has a zero modes. From the analysis of the previous
sections we know that indeed the operator (\ref{o1}) has a unique
zero eigenvalue. This eigenvalue corresponds to a massless photon
in four-dimensions which is the unique bound state of the
attractive delta-function potential in eq.(\ref{pot}). On top of
this, there exist a tower of massive KK states given in table
(\ref{t1}).  On the other hand, the operator of eq.(\ref{o2}) does
not have a zero mode. As a result, there is no bound state, no
massless scalars thereof and all four-dimensional scalars coming
from the components of $n+4$-dimensional vector appear massive.

\section{Exponentially large extra dimensions}

It has recently be proposed that the hierarchy problem, the
unnatural smallness of the ration $m_{EW}/M_P$ of the electroweak
scale $m_{EW}\sim 10^3$ {\rm GeV} to the four-dimensional Planck
scale $M_P\sim 10^{18}$ {\rm GeV} can be resolved in a higher
$n\!+\!4$-dimensional setting. In such a framework, $M_P$ is
related to the $n\!+\!4$-dimensional Planck scale $M_{P(n+4)}$ by
\be
M_P^2=M_{P(n+4)}^{n+2} V(X^n)\, , \ee where $V(X^n)$ is the volume
of the internal space $X^n$.   Usually, for a more or less
isotropic space $X^n$ of characteristic scale $R$ we have
\be
V(X^n)= \alpha R^n\, , \label{V} \ee where $\alpha$ is an
$R$-independent constant so that,
\be
M_P^2=M_{P(n+4)}^{n+2} R^n\, . \ee Thus, as have been pointed out
in \cite{ADD}, taking $M_{P(n+4)}\sim m_{EW}$, the hierarchy
problem is solved if the scale of the internal space $X^n$ is
large. For two extra dimensions for example with $m_{EW}\sim 10^3$
{\rm GeV} we get $R\sim 1$mm. However, in such a scenario one
expects that $R\sim M_{P(n+4)}^{-1}$ as this is the scale in the
higher dimensional theory. Thus,  the hierarchy $m_{EW} R$ ($\sim
10^{15}$ for two extra dimensions) has still to be explained. This
hierarchy can be traced back to eq.(\ref{V}), and as we will see
here  this is not in general the case. Namely, we will construct a
vacuum configuration in which the volume of the internal space
$V(X^n)$ although by dimensional reasons satisfy eq.(\ref{V}), the
constant $\alpha$ is an exponential function of $R$. In this case,
even $R\sim M_{P(n+4)}^{-1}$, an exponentially large volume
emerges so that no hierarchy $m_{EW} R$ appears.

We will consider again the metric (\ref{met1}) where now
$\Lambda(z)$ is
\be
\L(z)=-\mu_1|z+L|+\mu_2|z-L| \, , \ee where $\mu_1,\mu_2$ are, as
before, dimensionful constants and $2L$ is the distance between
the two branes sited at $-L,L$. It is natural to assume that all
scales, namely, $\mu_1,\mu_2,1/a,1/L$ are of the order of the
six-dimensional Planck scale $M_{P(6)}$.
 The volume of the transverse space is finite if $\mu_1>\mu_2$ and
 in this case we find
\be
V={2\pi a \mu_1 \over {\mu_1^2-\mu_2^2}}e^{4\mu_2L}+{2\pi a \mu_2
\over {\mu_1^2-\mu_2^2}}e^{-4\mu_1L}\, . \ee The first term in the
expression above dominates and the four-dimensional Planck scale
$M_P$ is then
\be
M_P^2=M_{P(6)}^4V=M_{P(6)}^4{2\pi a \mu_1 \over
{\mu_1^2-\mu_2^2}}e^{4\mu_2L}\, . \ee For $\mu_1=2$\,{\rm TeV}
,$\mu_2=1${\rm TeV}, $1/a=1$\,{\rm TeV},   the value $M_P\sim
10^{18}$ {\rm GeV} for the four-dimensional Planck scale  is
obtained for $1/L=15$\, {\rm TeV}. The masses of the KK states are
now at the {\rm TeV} scale as was originally proposed in \cite{IA}

\vspace{.6cm}

\noindent {\bf Acknowledgement}:  The author has benefited by
discussions with A. Brandhuber, E. Floratos, A. Riotto and G.
Tiktopoulos. This work is supported by a GGET grant No.
$97E\L/71$.

%



\end{document}


\begin{thebibliography}{99}

\bibitem{LB} J.D. Lykken, Phys. Rev.  D 54 (1996) 3693,
hep-th/9603133;\\
 C. Bachas, 1995 (unpublished) and  JHEP  9811 (1998) 023,
hep-th/9807415.
\bibitem{IA}I. Antoniadis, Phys. Lett. B 246 (1990)377.
\bibitem{W} E. Witten, Nucl. Phys. B 471 (1996)135, hep-th/9602070;
J.D. Lykken, Phys. Rev. D 54 (1996)3693, hep-th/9603133.

\bibitem{ADD} N. Arkani-Hamed, S. Dimopoulos and G. Dvali, Phys. Lett. B
429 (1998)263, hep-th/9803315; I. Antoniadis, N. Arkani-Hamed, S.
Dimopoulos and G. Dvali, Phys. Lett. B 436 (1998)257,
hep-th/9804398. \bibitem{RS} T. Regge and C. Teitelboim, MArcel
Grossman Meeting on GR, Trieste 1975, North Holland;\\ V. Rubakov
and M. Shaposhnikov, Phys. Lett. B 125 (1983) 136.
\bibitem{hw} P. Ho\v rava and E. Witten, Nucl. Phys. {\bf B460}
(1996) 506, hep-th/9510209.

\bibitem{wit} E. Witten, Nucl.Phys.{\bf B471} (1996) 135;
hep-th/9602070.



\bibitem{TV} T.R. Taylor and G. Veneziano, Phys. Lett. B 212
(1988)147.
\bibitem{DDG} K.R. Dienes, E. Dudas and T. Gherghetta, Phys. Lett.
B 436 (1998)55, hep-ph/9803466; Nucl. Phys. B 537 (1999) 47,
hep-ph/9806292.
\bibitem{NY} P. Nath and M. Yamaguchi, Phys. Rev. D 60 (1999)
116006,hep-ph/9903298; T.G. Rizzo and J.D. Wells, hep-ph/9906234;
\\R. Casalbuoni , S. De Curtis, D. Dominici and R. Gatto,
hep-ph/9908299; \\A. Delgado, A. Pomarol and M. Quiros,
hep-ph/9911252.
\bibitem{RS1}L. Randall and R. Sundrum, Phys. Rev. Lett. 83 (1999) 3370,
hep-ph/9905221; hep-th/9906064.
\bibitem{CK}  A.G. Cohen and  D.B. Kaplan, hep-th/9910132.
 \bibitem{PR} H. Davoudiasl, J.L. Hewett and T.G. Rizzo,
 hep-ph/9911262;\\
 A. Pomarol, hep-ph/9911294.
\bibitem{GZ} M. Gell-Mann and B. Zwiebach, Nucl. Phys. B 260 (1985)
569; Phys. Lett. B 147 (1984) 111. \\
 H. Nicolai and C. Wetterich, Phys. Lett. B 150 (1985) 347.
\bibitem{wet} C. Wetterich, Nucl. Phys. B 242 (1984) 47, Nucl.
Phys. B 244 (1984) 359, Nucl. Phys. B 253 (1985) 366.
\bibitem{WWW} E. Witten, Proc. 1983 Shelter Island II Conf., 1983.
427.
\bibitem{KK} A. Kehagias, hep-th/9906204.
\bibitem{Nwt} S. Dimopoulos and G. Giudice, Phys. Lett. B 379 (1996)
 105, hep-ph/9602350;\\
 J. C. Long, H.W. Chan and J.C. Price, Nucl. Phys. B 539 (1999) 23,
 hep-ph/9805217;\\
 N. Arkani-Hamed, S. Dimopoulos and  G. Dvali, Phys. Rev. D 59 (1999) 086004,
  hep-ph/9807344;\\
  E. Floratos and G. Leontaris, hep-ph/9906238.
\bibitem{KS} A. Kehagias and  K. Sfetsos,  hep-ph/9905417.





\end{thebibliography}

\begin{thebibliography}{99}


\bibitem{ADD} N. Arkani-Hamed, S. Dimopoulos and G. Dvali, Phys. Lett. B
429 (1998)263, hep-th/9803315; I. Antoniadis, N. Arkani-Hamed,
S. Dimopoulos and G. Dvali, Phys. Lett. B
436 (1998)257, hep-th/9804398.
\bibitem{IA}I. Antoniadis, Phys. Lett. B 246 (1990)377.
\bibitem{W} E. Witten, Nucl. Phys. B 471 (1996)135, hep-th/9602070;
J.D. Lykken, Phys. Rev. D 54 (1996)3693, hep-th/9603133.
\bibitem{RS} L. Randall and R. Sundrum, {\it A large mass hierarchy from a
small extra dimension}, hep-ph/9905221; {\it An alternative to
compactification}, hep-th/9906064.
\bibitem{ZZ} L. Girardello, M. Petrini, M. Porrati and
 A. Zaffaroni,  JHEP 9812:022 1998, hep-th/9810126.
\bibitem{KPW} A. Khavaev, K. Pilch and N.P. Warner, {\it New vacua of gauged
N=8 supergarvity in five dimensions}, hep-th/9812035.
\bibitem{FGPW} D.Z. Freedman, S.S. Gubser,  K. Pilch and N.P. Warner,
{\it Renormalization group flows
from holography-supersymmetry and a c-theorem}, hep-th/9904017.
\bibitem{Mald}J. Maldacena, Adv. Theor. Math. Phys. {\bf 2} (1998) 231,
hep-th/9711200
\bibitem{KP} S.S. Gubser, I.R. Klebanov and  A.M. Polyakov,
 Phys. Lett. {\bf B428} (1998)105, hep-th/9802109.
\bibitem{WW}E. Witten, Adv. Theor. Math. Phys. {\bf 2} (1998) 253,
hep-th/9802150.
\bibitem{NW}B. de Wit, H. Nicolai and N.P. Warner, Nucl. Phys. B 255 (1984) 29;
 B. de Wit and H. Nicolai, Nucl. Phys. B 281 (1987) 211.
\bibitem{LS1} A. Lukas, B.A. Ovrut, K.S. Stelle and D. Waldram,
Phys. Rev. D 59 (1999) 086001, hep-th/9803235.
\bibitem{N} W. Nahm, Nucl. Phys. B 135 (1978)149.
\bibitem{CR} E. Cremmer, J. Scherk and J. Schwarz, Phys. Lett. B 84 (1979)83;
E. Cremmer, in {\it Superspace and supergravity}, ed. S.W. Hawking and M.
Ro$\check{c}$ek, Cambridge Univ. Press, 1981.
\bibitem{GRW} M. G\"unaydin, L.J. Romans and N.P. Warner, Phys. Lett. B 154
(1985)268.
\bibitem{GRW1} M. G\"unaydin, L.J. Romans and N.P. Warner, Nucl. Phys. B 272
(1986)598.
\bibitem{GRW2}   M. G\"unaydin, L.J. Romans and N.P. Warner, Nucl. Phys. B
272 (1986) 598.
\bibitem{GT} C.W. Gibbons and P.K. Townsend, Phys. Rev. Lett. 71 (1993)3754.
\bibitem{T} H. L\"u, C.N. Pope, S. Sezgin and K.S. Stelle, Nucl. Phys. B
456 (1995) 669, hep-th/9508042; Phys. Lett. B 371 (1996) 46, hep-th/9511203.
\bibitem{LPT} H. Lu, C.N. Pope and P.K. Townsend, Phys. Lett. B 391 (1997)39,
hep-th/9607164.
\bibitem{T1} P.M. Cowdall, H. L\"u, C.N. Pope, K.S. Stelle and P.K.
Townsend, Nucl. Phys. B 486 (1997) 49,  hep-th/9608173.
\bibitem{GR} G.W. Gibbons and P. Rychenkova, {\it Single-sided domain walls
in M-theory}, hep-th/9811045.
\bibitem{C} M. Cvetic, J.T. Liu, H. L\"u and C.N. Pope, {\it Domain wall
supergravities from sphere reduction},  hep-th/9905096.
\bibitem{MS} L.J. Romans, Phys. Lett. B 169 (1986)374
\bibitem{BRE} E. Bergshoeff, M. de Roo and
E. Eyras, Phys. Lett. B 413 (1997)70, hep-th/9707130
\bibitem{MS1}  N. Kaloper and R.C. Myers, JHEP 9905 (1999)010, hep-th/9901045.
\bibitem{V} H. Verlinde, {\it Holography and compactification}, hep-th/9906182.
\end{thebibliography}
\end{document}